\documentclass[12pt]{iopart}

\usepackage{graphicx}
\usepackage[T1]{fontenc}
\usepackage[utf8]{inputenc}

\begin{document}

\title[Development and decay of vortex flows in viscoelastic fluids
  between concentric cylinders] {Development and decay of vortex flows
  in viscoelastic fluids between concentric cylinders}

\author{Renzo Guido, Felipe Rinderknecht, Cecilia Cabeza, Arturo
  C. Martí, Gustavo Sarasúa}

\address{Igu\'a 4225, Instituto de Física, Facultad de Ciencias,
  Universidad de la República, Montevideo, Uruguay}

\ead{marti@fisica.edu.uy}

\begin{abstract}
We study the development and decay of vortex in viscoelastic fluids
between coaxial cylinders by means of experiments with solutions of
polyacrylamide and glycerin and numerical simulations. The transient
process is triggered when the inner cylinder is either abruptly
started or stopped while the outer is kept fixed.  The azimuthal
velocity, obtained by means of digital particle velocimetry, exhibits
oscillations before reaching the stationary state.  The development of
the vortex is characterized by means of the overshoot, i.e.  the
difference between the maximum and the stationary
velocity. Analogously, in the decay of the vortex, the azimuthal
velocity changes its direction and the relevant parameter is the
undershoot defined as the maximum reversed transient velocity. To get
a deeper insight into this phenomenon, the experimental results are
supplemented with numerical simulations of rheological models as the
Oldroyd-B and White-Metzer. The results obtained with the first model
reveal the dependence of the overshoot and undershoot with the
elasticity number of the fluid.  Using the White-Metzer model we
explain the increase of the overshoot produced by the reduction of the
solvent viscosity in terms of the shear-thinning effects.
\end{abstract}

\maketitle

\section{Introduction}
Many industrial applications, as the production of food, lubricants,
paints, among many others, involve the flow of non-Newtonian fluids,
in general much more complex than their Newtonian counterpart
\cite{irgens}.  Non-Newtonian flows have been studied in numerous
configurations, however, one aspect that has received comparatively
less attention is their transient dynamics \cite{Joo,Shaqfeh}. Here we
consider the development and decay of vortex flows in non-Newtonian
fluids \cite{Qi,Fetecau2008,Khan2008}. In particular, we consider a
system composed of two cylinders in which the outer is kept fixed
while the inner can be abruptly put in motion or stopped.  Due to the
presence of the ``hoop" stress in flows of viscoelastic fluids with
curved streamlines \cite{Joo}, we expect the behavior of the transient
rotational flows in such systems exhibits remarkable differences with
that developed in Newtonian fluids.
 
Several experimental studies have been conducted to study the
development and decay of vortex flow of viscoelastic fluids in
cylindrical containers \cite{Hill,Day,Escudier,Tamano}. However, the
experimental study of the transient vortex flow in viscoelastic fluids
between cylinders has received little attention.  In a pioneering
contribution, Groisman and Steinberg \cite{jos} reported the influence
of small polymer addition on the stability and pattern selection of
the Taylor-Couette flow. In contrast with the well-known Newtonian
case, in the polymeric flow they found oscillatory behavior attributed
to the fluid elasticity. Later, the flow of different polymer
solutions between coaxial cylinders was also experimentally considered
in \cite{Borja}, where the variation of the torque on the inner
cylinder compared with the Newtonian case was obtained and showed to
exhibit hysteretic behaviors.

The unsteady rotational flows of viscoelastic fluids have been
theoretically studied making use of different models. Using a
fractional Maxwell model, Qi and Jin \cite{Qi} analyzed two unsteady
flows between coaxial cylinders. In the first the outer cylinder
performs a simple harmonic motion while the inner cylinder is kept
still while in the second the outer cylinder suddenly starts up and
rotates at a constant speed while the inner cylinder is kept
stationary. In these problems the theoretical model predicted either
oscillations or an initial peak followed by a relaxation to an steady
value respectively in the evolution of the azimuthal velocity.
Fractional calculus was proposed to obtain exact solutions using for
the decay of a potential vortex in Oldroyd-B and Maxwell fluids
\cite{Fetecau2008,Khan2008}.  Closed expressions for the velocity
components and their dependence with the model parameters and the
fractional coefficients were reported in these studies.

In this work we consider two complementary transient dynamics. In the
first one we explore the start up of the swirling flow from the
quiescent fluid after the inner cylinder suddenly begins to rotate. In
the second one, we examine the decay of the initially steady vortex
flow after the inner cylinder becomes abruptly at rest.  The
experimental results show that, in contrast with Newtonian flows, the
azimuthal velocity reaches a maximum before decaying in the case of
the development of a vortex, while it changes its direction in the
case of the decay.  These behaviors can be characterized by two
dimensionless parameters, denoted overshoot and undershoot. In order
to get a deeper insight into this phenomenon we carried out numerical
simulations using the Oldroyd-B and White-Metzer (WM) models. These
simulations allow to identity the elasticity of the fluid and the
shear-thinning effects as being related to this peculiar transient
behavior. The rest of this article is organized as follows. In section
2 we describe the experimental setup, the methodology and the working
fluids. In section 3 we present the experimental results.  The
numerical results described in Section 4 generalize the experimental
results.  Finally, Section 5 presents the discussion and conclusion.

\section{Experimental setup}

\begin{figure} 
\centering
\includegraphics[width=0.48\textwidth]{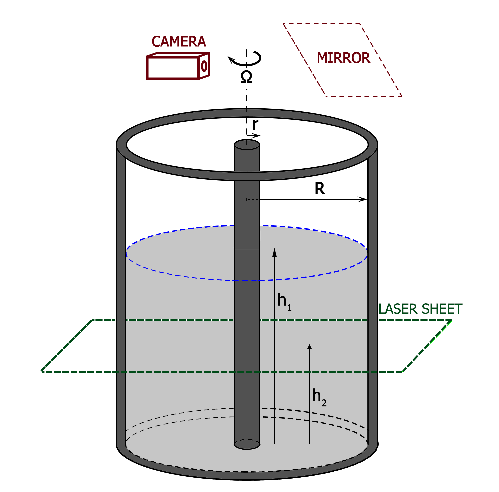}
\includegraphics[width=0.48\textwidth]{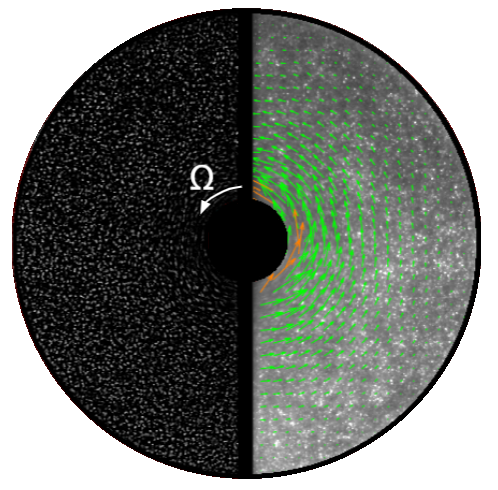}
\caption{Panel (a) schematizes  the experimental set
 up. The cylinders are shown in solid lines, with radius $R$ and
  $r$. The laser plane is indicated at a height $h_2$ with a dash
  line, and the fluid's surface is at height $h_1$. The camera is
  $1.5~m$ from the axis and points to the mirror above the
  cylinders. Panel (b) is a example of a picture taken from the video
  (left) and its DPIV (right). The height of the plane is $45$ mm. The
  camera only recorded a half of the circle, so the picture on the
  left was mirrored. }
\label{fig:Apparatus}
\end{figure}

The core of the experimental setup schematized in
Fig.~\ref{fig:Apparatus} consists of two concentric acrylic
cylinders. The external of radius $R$ = 41.65 $\pm$ 0.05 mm is kept
fixed, and the internal whose radius is $r$ = 4.75 $\pm$ 0.05 mm is
able to rotate driven by a direct current motor. Its maximum angular
velocity is $\Omega=$ 7.0 rad/s. The height of the fluid column is
$h_{1}$ = 100 $\pm$ 1 mm. The bottom is fixed and the fluid upper
surface is free at atmospheric pressure.

The fluids used in this experiment were four shear-thinning solutions
of Polycarylamide (92560-50G Sigma-Aldrich) in mixtures of water and
glycerin enumerated in Table \ref{TablaReologia}. The rheological
properties of the different fluids were measured at $20^{\circ}C$,
with an Anton Paar Physica MCR 301 rheometer.  For each fluid, the
strain dependent dynamics viscosity, $\eta( \dot{\gamma})$, was fitted
to the Carreau model $\eta(\dot{\gamma})=\eta_0 (1+(\lambda
\dot{\gamma})^2)^{(n-1)/2}$ to obtain the viscosity at zero-shear
rate, $\eta_0$, and the characteristic time, $\lambda$, as shown in
this table.

The relevant dimensionless parameters to characterize the viscoelastic
effects are the Deborah and elasticity numbers. The Deborah number is
defined as De = $\lambda U /R$, where $R$ is the radius of the inner
cylinder, $U$ is the velocity at the surface of that cylinder and
$\lambda$ is the characteristic time. The elasticity number is defined
as E = De/Re, where Re = $ U R /\nu_0$ is the Reynolds number and
$\nu_0$ the zero-shear kinematic viscosity. Using the angular velocity
of the inner cylinder $\Omega$, these numbers can be written as
$\textrm{De} = \lambda \Omega,$ $\mathrm{Re} = \Omega R^2 /\nu_0 $ and
$\mathrm{E}=\lambda \nu_0/R^2.$ The specific values for each fluid,
which are independent on $\Omega$, are listed in Table
\ref{TablaReologia}.

The velocity fields were measured by means of Digital Particle Imaging
Velocimetry (DPIV), which is based on 2D cross-correlation of
consecutive images.  To this end, the fluid was seeded with neutrally
buoyant nearly-spherical polyamide particles (Dantec Dynamics,
Denmark). The particles with mean diameter of 50 $\mu m$ and Stokes
number of approximately $10^{-4}$ can be assumed to closely follow the
flow. The particles were illuminated in the horizontal plane shown in
the figure at height $h_{2}$=45~mm from the bottom. A green laser
LaserGlow (model LSR-0532-PFH-005500-05n) was used for this
purpose. The light reflected by the particles was captured by a
Pixelink (model PL-B7760) digital camera using a mirror at
45\textdegree \, fixed above the cylinders and working 45 fps. As the
velocity in the region close to the external cylinder is small, the
DPIV was performed averaging three consecutive frames.  Then, the
velocity field was obtained at a sampling frequency of 15 Hz in a
Cartesian grid.  At the rotation velocities involved in our
experiments the velocity fields observed were almost axysymmetric. To
obtain the azimuthal component each grid point was classified
according to its distance to the rotation axis in rings of $2.05$ mm
thickness. At each time, we calculated the averaged azimuthal velocity
and its standard deviation.

\begin{table}
\centering
 \begin{tabular}{||c || c c c c ||} 
 \hline
 Properties &  P1G30 & P1G60 & P2G30 & P2G60 \\ [0.5ex] 
 \hline \hline
 Glycerin (\%) & $30$ & $60$ & $30$ & $60$ \\ 
 \hline
 Polyacrylamide (\%) & $1.0$ & $1.0$ & $2.0$ & $2.0$ \\ 
 \hline \hline
 $\mu_0$(Pa$\cdot $s) & $4.65$ & $4.40$ & $107$ & $157$ \\ 
 \hline
 $\nu_0( \times 10^{-3}$ m$^2$/s) & $4.39$ & $3.93$ & $101$ & $140$ \\  \hline
$\lambda$(s)  & $9.8$ & $2.2$ & $42$ & $46$ \\ 
 \hline
 $\mathrm{Re}(\times 10^{-3})$  & $29.0$ & $32.5$ & $0.91$ & $1.26$ \\ 
 \hline
$\mathrm{De}$ & $55.4$ & $12.4$ & $237$ & $260$ \\ 
 \hline
 $\mathrm{E} (\times 10^{3})$ & $1.91$ & $0.383$ & $188$ & $286$ \\ [1ex]
 \hline
\end{tabular}
\caption{Fluid samples used along the experiments and their
  rheological properties. The rotational velocity used for calculating
  the Reynolds and Deborah number was 5.65 rad/s.}
\label{TablaReologia}
\end{table}

\section{Experimental results and analysis}
\subsection{Start-up of the vortex flow}

\begin{figure}\centering
\includegraphics[width=0.49\textwidth]{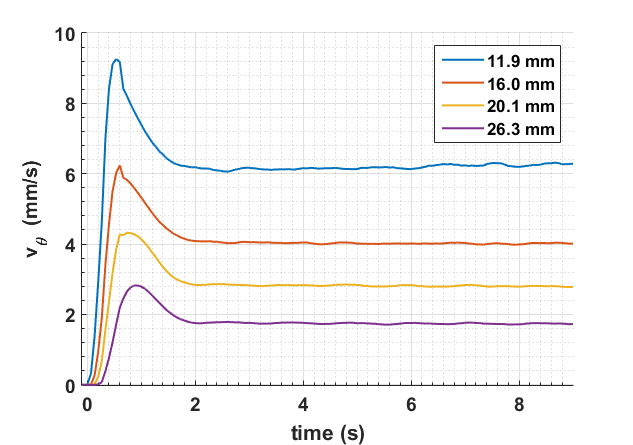}
\includegraphics[width=0.49\textwidth]{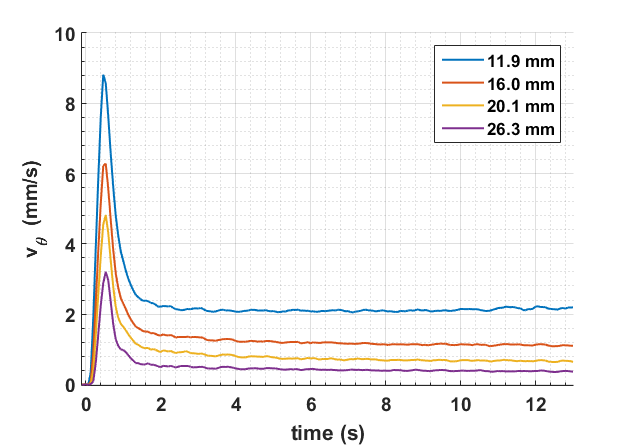}
\caption{Temporal evolution of the azimuthal velocity at different
  distances to rotation center indicated in the legend boxes.  Panel
  (a) corresponds to the fluid P1G30 and (b) to the P2G30. The initial
  time, $t=0$, corresponds to the instant when the motor was turned
  on. The uncertainties are nearly constant and their mean values are
  $0.4$ mm/s, $0.2$ mm/s, $0.2$ mm/s and $0.1$ mm/s respectively for
  each axial distance.}     \label{fig:TurnOn}
\end{figure}

In this section we describe the experiments performed to study the
development of the vortex flow departing from the quiescent fluid. In
these experiments the inner cylinder is initially at rest and abruptly
begins to rotate with a constant angular velocity $\Omega$.  In
Fig.~\ref{fig:TurnOn} we show the temporal evolution of the azimuthal
component of the velocity at different axial distances for the fluids
P1G30 and P2G30. In all the cases, the azimuthal velocity reaches a
maximum and then decreases approaching a stationary value. This
characteristic behavior is present in all the fluids analyzed.

 Inspired in control theory \cite{katsuhiko2010modern}, to quantify
 this characteristic behavior we define the overshoot as the
 dimensionless parameter $M_p = (v_{max} - v_{st})/{v_{st}} $ where
 $v_{max}$ and $v_{st}$ are the maximum and the stationary value of
 the velocity.

\begin{figure}
\begin{center}
\includegraphics[width=.8\textwidth]{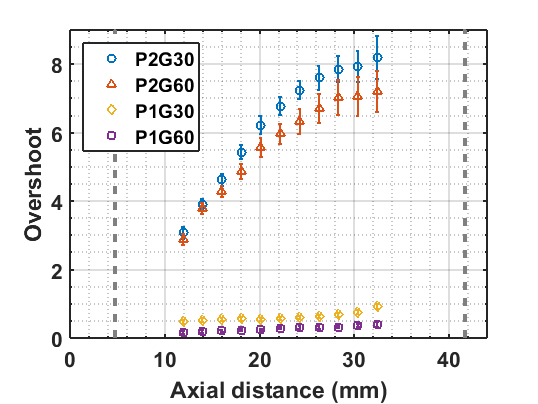}
\caption{Overshoot increases with the axial distance for all the
  samples.  The difference between the samples with 1\% of
  polyacrylamide and those with 2\% is enormous. The borders of the
  cylinders are indicated with vertical dashed lines.}
\label{fig:Overshoot}
\end{center}
\end{figure}

 In Fig.~\ref{fig:Overshoot} we plot the overshoot as a function of
 the distance to the center of rotation, $r$, for all the fluids
 samples considered. Several relevant aspects can be noted in this
 figure. First, we observe that in all the cases the overshoot
 increases with the radial coordinate indicating a relationship with
 positive streamlines curvature. Nonetheless, this effect is more
 notorious when the polymer concentration is 2\%, fluids P2G30 and
 P2G60, in comparison with the samples in which the concentration is
 1\%, P1G30 and P1G60. This results clearly suggest a positive
 relationship of the overshoot with the elastic modulus.  In addition,
 when we compare fluid samples with the same polymer concentration but
 different glycerin concentration we note that the overshoot increases
 with decreasing the concentration suggesting that the viscous part of
 the fluid also plays an important role in this phenomenon.

\begin{figure}
\begin{center}
\includegraphics[width=.8\textwidth]{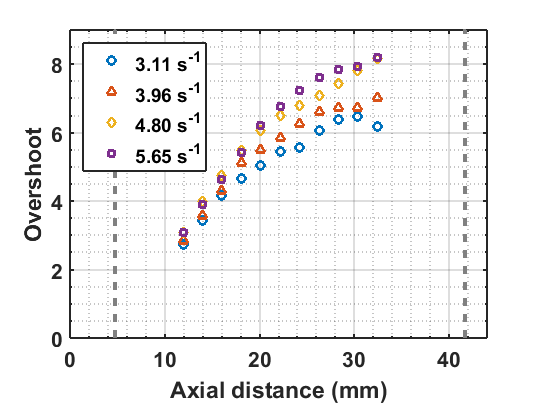}
\caption{Overshoot as a function of axial distance for four rotational
  velocities indicated in the legend box. The fluid sample is the
  P2G30. As in the previous figure, the borders of the cylinders are
  indicated with vertical dashed lines.}
\label{fig:P2G30VelocidadOS}
\end{center}
\end{figure}

After that we repeated the experiments starting again from a fluid at
rest but considering different final angular velocities.  In
Fig.~\ref{fig:P2G30VelocidadOS}, corresponding to fluid sample P2G30,
we observe the overshoot as a function of the axial distance for four
different rotational velocities. We notice a dependence in which the
overshoot increases with increasing the angular velocity. Nonetheless,
this effect is more noticeable for the regions close to the inner
cylinder.

\subsection{Decaying vortex}

\begin{figure}
     \centering
         \includegraphics[width=0.49\textwidth]{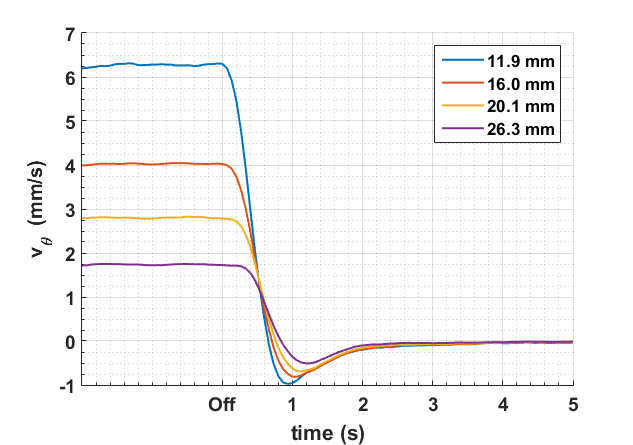}
         \includegraphics[width=0.49\textwidth]{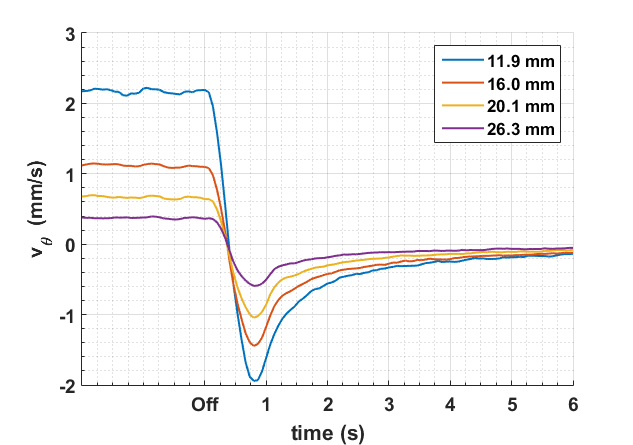}
  \caption{The undershoot is manifested in the temporal evolution of
    the azimuthal velocity when the inner cylinder is abruptly stopped
    at the time indicated in the horizontal axis. The different curves
    correspond to different axial distances indicated in the legend
    boxes.  Panels (a) and (b) correspond to P1G30 and P2G30
    respectively.  The uncertainties in the velocities are 6\%.}
        \label{fig:TurnOff}
\end{figure}

In this section we study the decaying of the vortex flow when the
inner cylinder, initially rotating at constant speed, is abruptly
stopped. As shown in Fig.~\ref{fig:TurnOff}(a), corresponding to he
fluid P1G30, the azimuthal velocity starts at a steady value, then it
decreases quickly, reverses its direction until reaching a maximum
negative (opposite) value and then decreases (in magnitude)
approaching a null velocity. This behaviour is presented for all the
curves in Fig.~\ref{fig:TurnOff}(a), corresponding to different
distances from the cylinder's axis.  The panel (b) shows the temporal
evolution of the azimuthal velocity for the fluid P2G30, where the
reversal of the velocities is also observed. Similar to that observed
in the development of the vortex, the effect is more notorious in the
P2G30 fluid than for the P1G30 in which the concentration of
polyacrylamide is larger in the first than in the latter.

\begin{figure}
\begin{center}
\includegraphics[width=.8\textwidth]{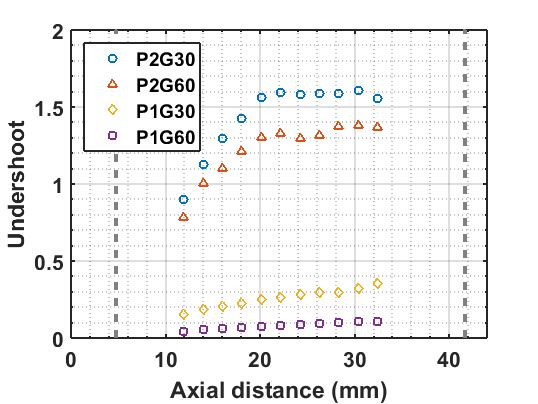}
\caption{Undershoot as a function of axial distance for all the
  fluid sample. The difference observed between high and small
  polyacrilamide concentration is remarkable.  At small elastic modulus,
  the undershoot increases with axial distances. Also, the smaller
  solvent viscosity fluid P1G30 exhibits a larger reversal. At high
  elastic modulus, the effect increases with axial distance until it
  reaches a plateau. As in the previous experiment, the samples P2G60
  and P2G30 reveal similiar behaviours.}
\label{fig:Undershoot}
\end{center}
\end{figure}

Similarly to the overshoot defined in the development of the vortex,
it is possible to introduce a dimensionless parameter for the reversal
of the fluid, referred to as undershoot or relative maximum negative
peak value \cite{katsuhiko2010modern}, $M_{mn} =
|\frac{v_{mn}}{v_{st}}|$ where $v_{mn}$ is the maximum negative (or
opposite) velocity. The dependence of the undershoot with the axial
distance for each fluid is shown in Fig.~\ref{fig:Undershoot} for all
the fluid samples studied.  It can be noticed that samples with higher
elastic components exhibit notoriously larger undershoots, a
qualitative behaviour similar to that reported in
Fig. \ref{fig:Overshoot} for the overshoot. Moreover, comparing
samples with the same polyacrylamide concentration and different
viscosity, those with smaller viscosity exhibit a larger effect.

\section{Numerical Simulations}

\subsection{Oldroyd-B governing equations}
To get a deeper insight into the transient behavior of the vortex
flows we performed numerical simulations using the well-known
Oldroyd-B model \cite{oldroyd1950formulation} which can be considered
as an extension of the Maxwell model and is equivalent to a fluid
filled with elastic beads and spring dumbbells.  Our aim was to
determine the role of the elastic component of the fluid and the
solvent viscosity in the overshoot and undershoot observed in the
experiments reported in the previous section. According to the
Oldroyd-B model, the dimensionless governing equations of a
viscoelastic incompressible fluid are given by
\begin{equation}
\mathrm{Re} \frac{D {\bf v}}{Dt}= -\nabla p + \beta \nabla^2 {\bf v}
+\nabla \cdot {\bf \sigma},
\label{ob1}
\end{equation}
where ${\bf \sigma}$ is the polymeric stress tensor, $\beta$ is the
ratio between the solvent and total viscosities, $\beta=
\mu_s/(\mu_s+\mu_p)$, being $\mu_p$ the polymer viscosity.  The
polymeric stress tensor of an Oldroyd-B model satisfies the evolution
equation
\begin{equation}
\mathrm{De} \ ( \frac{\partial {\bf \sigma}}{\partial t} + {\bf v}
\cdot \nabla {\bf \sigma} \ - {\bf \sigma} \cdot \nabla{\bf v} -
\nabla {\bf v}^\dag \cdot {\bf \sigma} \ ) + {\bf \sigma} = -
(1-\beta) ( \nabla {\bf v} + \nabla {\bf v}^\dag).
\label{ob2}
\end{equation}
In the statistical derivation of the Oldroyd-B model starting from a
microscopic approach it can be shown that
\begin{equation}
\lambda = \frac{3 \pi a \mu_s   }{4 k_B T \beta_L^2},
\label{la}
\end{equation}
\begin{equation}
\mu_p = \frac{3 \pi m a \mu_s   }{4 \beta_L^2}
\label{mup}
\end{equation}
where $k_B$ is the Boltzmann constant, $T$ the absolute temperature,
$m$ the polymer concentration, and, finally, $a$ and $\beta_L^{-1}$
are characteristic lengths of the polymeric chain model
\cite{larson2013constitutive}.  Another important parameter for our
purposes is $G=\mu_p /\lambda =mk_B T$, which measures the influence
of the polymeric concentration on the fluid properties.

We solved the constitutive equations, Eqs.~(\ref{ob1}-\ref{ob2}),
using the finite element com\-pu\-ta\-tional package {\small COMSOL}
\cite{comsol1} which has been shown to be a reliable tool in different
computational fluid dynamics problems (see for example
Refs. \cite{Martell, Zhu}). The code validation has been done
considering the flow of a viscoelastic fluid past a cylinder. Assuming
axial symmetry, we performed simulations on two dimensional meshes in
which the inner and outer cylinders are located at $R$ and $10R$
respectively and the fluid depth is $h= 20 R$. A more refined mesh was
used near the inner cylinder to avoid numerical
instabilities. Neglecting air viscosity and surface tension, the
interface between the viscoelastic fluid and the air was simulated as
an horizontal wall with a slip boundary condition. This approximation
agrees very well with the experiments in which the interface only
slightly deviates from the horizontal plane.

\begin{figure}
\includegraphics[width=0.49\textwidth]{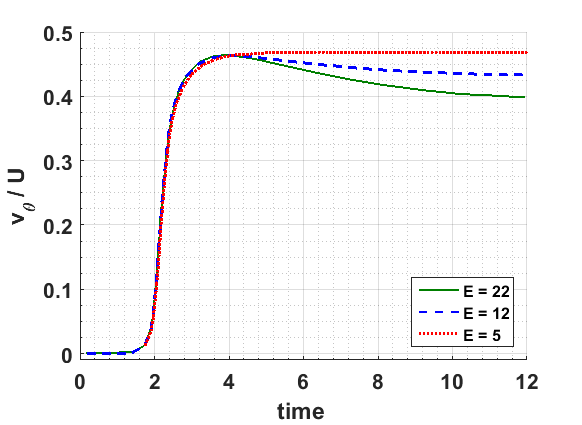}
\includegraphics[width=0.49\textwidth]{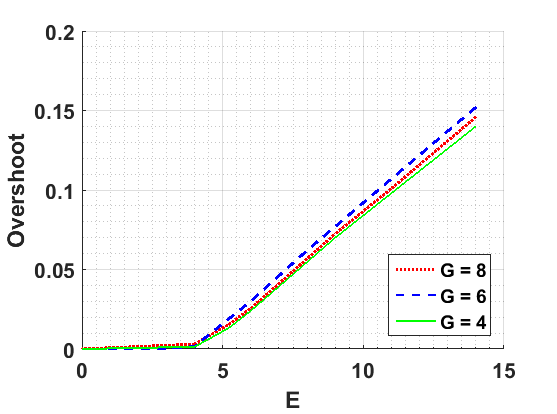}
\caption{Panel (a) shows  dimensionless azimuthal velocity as a function
  of time for  different elasticity values, E, indicated in the
  legend box. Panel (b) corresponds to the overshoot parameter $M_p$
  as a function of  E  and different
  values of the parameter G which characterises the polymeric
  concentration indicated. In both panels $Re=0.1$.}
    \label{fig:SimulacionesOldroyd}
\end{figure}

We numerically simulated the temporal evolution of the velocity field
starting with a fluid at rest for three different meshes with
$2^{13}$, $2^{14}$ and $2^{15}$ elements. The differences between the
three meshes were negligible, with a maximum difference in azimuthal
velocity of $0.6$ \% for $2^{13}$ and $2^{15}$ elements, and $0.3$ \%
for $2^{14}$ and $2^{15}$ elements. We defined dimensionless
quantities based on the stationary angular velocity and the inner
cylinder radius. In Fig. \ref{fig:SimulacionesOldroyd} the curves of
dimensionless azimuthal velocity $v_{\theta}/U$ at the position $r = 2
$, $z = 10$ are compared for different values of E. These curves
reveal that according to the Oldroyd-B model the overshoot is clearly
enhanced with the increment of E.
 
In addition, the numerical results indicate that there is always an
overshoot for any finite value of $\lambda$, but its value can be
strongly reduced as E is decreased. Figure
\ref{fig:SimulacionesOldroyd} shows the dependence of $S$ with E, for
Re=0.1 and different values of $G$. As can be seen from the figure,
the overshoot exhibits almost a linear dependence with E for values
above a characteristic value E$_s$, while for E $<$ E$_s$ the
overshoot values are very small. This behavior explains the large
differences in the undershoot values obtained for the distinct fluid
samples considered in the experiments.
 
\begin{figure}
\includegraphics[width=0.49\textwidth]{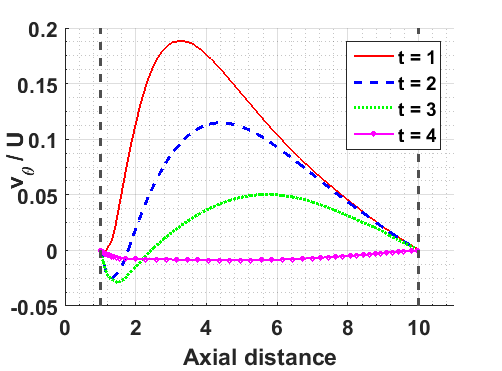}
\includegraphics[width=0.49\textwidth]{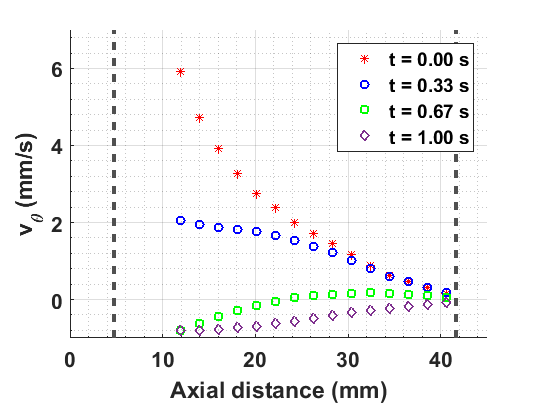}
\caption{Azimuthal velocity as a function of $r$ at the different
  times indicated in the legend box. In panel (a) corresponding to the
  numerical the simulation of the Oldroyd-B fluid the units of time
  are arbitrary. Panel (b) corresponds to experimental data for the
  P1G30 fluid.}
    \label{fig:Decaimiento}
\end{figure}

In Fig. \ref{fig:Decaimiento} we show the azimuthal velocity,
$v_\theta$, at different radial positions during the decay of the
vortex flow after the inner cylinder abruptly stops. As it can be
observed in this figure, the temporal evolution and spatial dependence
of azimuthal velocity is very similar to those observed in the
experiments. The velocity first changes sign near the inner cylinder
and later the sign change propagates to larger radial positions.

As a consequence, it can be concluded that using the Oldroyd-B model
is possible to identify the roles of the elasticity component and
solvent viscosity in the formation and decay of the vortex.  It is
worth noticing that in accordance with the experiments, the velocity
profiles do not exhibit the oscillations that are present in the
solutions of \cite{Fetecau2008} for the fractional Oldroyd-B
model. This fact is probably caused by the differences in the boundary
conditions taken by these authors, while in \cite{Fetecau2008} the
vortex is unbounded, in our experiments and simulations the vortex
flow is confined in a cylindrical region.  Another remarkable aspect
is that the Oldroyd-B model is not adequate to give account of the
influence of the viscosity solvent $\mu_s$ on the evolution of the
vortex. In this model, the reduction of $\mu_s$ at constant
concentration $m$ implies that the relaxation time $\lambda$ and
consequently De are decreased, although the parameter $G$ remains
unaltered. Such reduction of De at constant $G$ always produces the
decrease of the overshoot, as it is shown in
Fig. \ref{fig:SimulacionesOldroyd}. Thus the Oldroyd-B model cannot
explain the increase of the overshoot due to the reduction of $\mu_s$
observed in the experiments. By comparing the overshoot in P1G60 and
P1G30 reported in the experiments of Section 2, it can be seen that
the last exhibited a larger overshoot than the former in spite of
presenting a lower viscosity solvent.  In order to study whether the
shear thinning effects cause this kind of dependence, in the next
section we introduce a model with strain dependent viscosity.
 \begin{figure}
\begin{center}
\includegraphics[width=.6\textwidth]{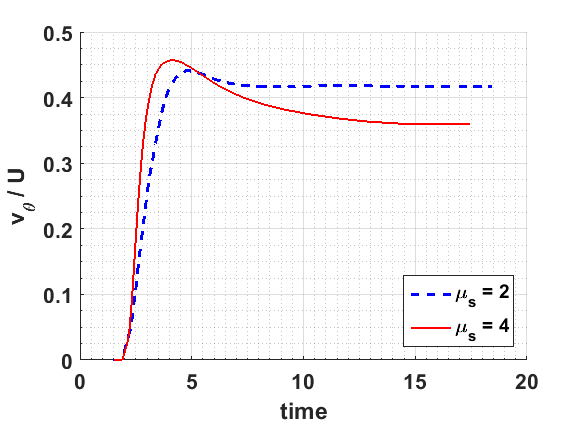}
\caption{Effect of the solvent viscosity variation in a simulation of
  the Oldroyd-B model.
  The solid line corresponds to De=0.5, Re = 0.025
  ($\mu_s=4$) and the dashed line to De=0.25, Re=0.05 ($\mu_s=2$).}
\label{fig:Od5}
\end{center}
\end{figure}

\subsection{Shear-thinning effects}

To study the influence of shear thinning effects on the evolution of
the vortex flow, we used the White-Metzer (WM) constitutive model
\cite{WM}.  The considered governing equations are formally the same
as Eqs. (\ref{ob1}), but with a relaxation time that is a function of
second invariant of the rate of deformation tensor ${\bf D}$,
$\dot{\gamma} =2 \textrm{tr}({\bf D}^2)$. Assuming that
$\lambda(\dot{\gamma})$ obeys the Carreau expression, due to the
relation $\mu_p=\lambda G$, it follows that the polymer viscosity is
also described by the Carreau model. The resulting total viscosity
$\mu=\mu_s+\mu_p$ is

\begin{equation}
\mu = \mu_{s} +\mu_{p0}[1+(\lambda_0 \dot{\gamma})^2]^{(n-1)/2} 
\end{equation}
where $\lambda_0$ and $\mu_{p0}$ are given by relations (\ref{la}) and
(\ref{mup}). Notice that when $\dot{\gamma}\rightarrow 0$, the
constant viscosity of the Oldroyd-B model is recovered.

\begin{figure}
\includegraphics[width=0.49\textwidth]{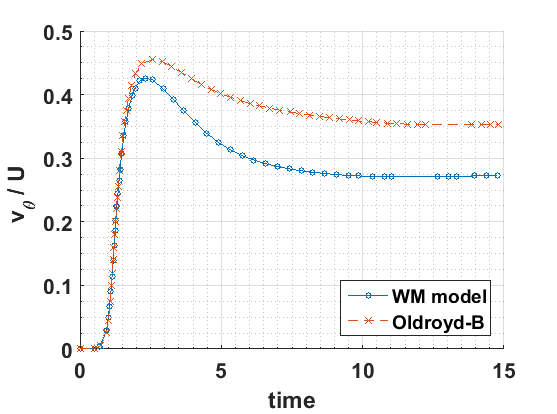}
\includegraphics[width=0.49\textwidth]{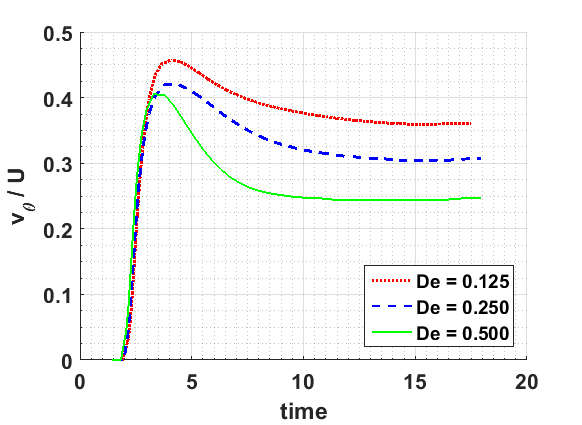}
\caption{Shear-thinning effects on the overshoot.  The temporal
  evolution of the azimuthal velocity at a fixed axial distance is
  plotted in panel (a) for the two models considered; Oldroyd-B (the
  overshoot results $M_p=0.28$) and WM model ($M_p=0.57$).  Panel (b)
  illustrates the angular velocity dependence on the overshoot using
  the WM model.  The Deberah number, De = $\lambda_0 \Omega$
  (indicated in the legend box) was varied by changing $\Omega$ and
  keeping $\lambda_0$ fixed.  In both cases E = 18, Re=0.1 and the
  power index in the WM model was $n=0.43$ (the value corresponding to
  P1G60 sample). The value of $\lambda_0$ used in the simulation of
  the WM model coincides with the one of $\lambda$ used in the
  simulation of the Oldroyd-B model. }
        \label{fig:CompararModelos}
\end{figure}

Figure \ref{fig:CompararModelos} presents the curves of azimuthal
velocity for the Oldroyd-B and WM models. The value of $\lambda_0$
used in the numerical simulations of the WM model coincides with the
value of $\lambda$ used in the simulations of the Oldroyd-B model and
both of them were performed with the same value of Re. These curves
show that the shear thinning effect favors the overshoot, which is
reflected in the increment of the parameter $M_p$.

Another fact observed experimentally is that the stationary angular
velocity of the inner cylinder clearly increases the overshoot. Figure
\ref{fig:CompararModelos}(b) shows the azimuthal velocities for three
different angular velocities according to the WM model. From these
curves it can be observed that the angular velocity of the inner
cylinder leads to an increment in the overshoot value. This result is
coherent with the influence of the shear thinning on this quantity
shown in panel (a). As the angular velocity is decreased, the strain
$\dot{\gamma}$ also decreases and we recover the Oldroyd-B model which
exhibits a smaller overshoot. On the opposite situation, when the
angular velocity is increased, the shear-thinning effects are more
notorious and the overshoot is enhanced.

The reported numerical results explain the effect of the variations in
the solvent viscosity. In general, an increment in the solvent
viscosity reduces the shear-thinning effects. Due to this reason, to
prepare a Boger fluid sample \textit{i.e.} a fluid with elastic
behavior but constant viscosity, a very high solvent viscosity is
recommended \cite{Bot,freire2019separation}. As a consequence, a
reduction in the solvent viscosity increases the shear-thinning
effects, which in turn increments the overshoot.

\section{Conclusion}
We studied the transient behaviour of viscoelastic vortex flows
between concentric cylinders by means of experiments and numerical
simulations. We study both the formation, that develops from the
quiescent fluid when the inner cylinder suddenly begins to rotate and
the decay, when the inner cylinder abruptly stops. In the experiments,
digital particle velocimetry was employed to obtain the velocity
fields in glycerin and polyacrylamide solutions while the numerical
simulations were based on implementations of the Oldroyd-B and
White-Metzer models using the commercial package COMSOL. For the
parameter values considered here the flows exhibit axial symmetry.

We found that, in contrast with Newtonian flows in which the azimuthal
velocity increases or decreases monotonically to a steady value, the
non-Newtonian cases are characterized by positive or negative maximum
values followed by the decay to a stationary value. As a way to
characterize these transient effects we introduced non-dimensional
parameters, overshoot and undershoot, similar to those well-known in
control theory.  The experimental results, supplemented with numerical
simulations reveal that, although this phenomenon is ubiquitous in
viscoelastic flows, their magnitude depend strongly on the distance to
the rotation center, the angular velocity and also on the fluid
characteristics.  In particular, we showed that the overshoot and the
undershoot depend on the elastic component and also on the solvent
viscosity.

The simulations performed with the Oldroyd-B model with boundary
conditions equivalent to experimental conditions reproduce very well
the observed flows, showing that this model is able to capture
qualitatively the phenomenon. In particular, the overshoot and
undershoot shapes agree very well. The several inflexion points in the
profile of the azimuthal velocity reported by Fetecau et al
\cite{Fetecau2008} were not observed suggesting that this disagreement
is probably caused by the fact the boundary conditions are different.
As a general observation the Oldroyd-B model can explain qualitatively
some aspects of the behavior of the unsteady vortex, but to describe
the dependence of the flow with the solvent viscosity and the
velocities at the material boundaries a model including shear thinning
effects is required.

As the Oldroyd-B model was not able to capture shear-thinning effects,
we introduced the WM model. Performing numerical simulations performed
with this model, we showed that the overshoot increment is produced by
the reduction of the solvent viscosity and it is directly related to
shear-thinning effects. The dependence of the overshoot with the
angular velocity of the inner cylinder was also explained with the WM
model. As a general result, the viscosity increase with the axial
distance due to the shear-thinning effect favors the overshoot. A
similar argument can be proposed to explain the similarities and
differences between the overshoot in Fig. \ref{fig:Overshoot} and the
undershoot in Fig. \ref{fig:Undershoot}. Both quantities grow with
axial distance until reaching a plateau where the value remains
constant. These increments with axial distance can be related to the
decrements of shear stresses also with axial distance and explain that
the undershoot reaches the stationary value closer to the center than
in the case of the overshoot.

The experimental results reported in our experiments, in particular,
the existence of the non-monotonic behavior of the azimuthal velocity
illustrates a notable characteristic of viscoelastic flows that has
previously received few attention. The academic and applied interest
in the dynamics of viscoelastic fluids encourages further experimental
and numerical investigations about this phenomenon.

\section*{Acknowledgements} We acknowledge financial support from grant
Fisica Nolineal (ID 722) Programa Grupos I + D CSIC 2018 (UdelaR, Uruguay)
and PEDECIBA (UdelaR, MEC, Uruguay).

\section*{References}

\bibliographystyle{IEEEtran}
\bibliography{/home/arturo/Dropbox/bibtex/mybib.bib}

\end{document}